\documentclass[a4paper, amsfonts, amssymb, amsmath, reprint, showkeys, nofootinbib, twoside]{revtex4-2}
\usepackage[english]{babel}
\usepackage[utf8]{inputenc}
\usepackage[colorinlistoftodos, color=green!40, prependcaption]{todonotes}
\usepackage{amsthm}
\usepackage{mathtools}
\usepackage{physics}
\usepackage{xcolor}
\usepackage{graphicx}
\usepackage[left=23mm,right=13mm,top=35mm,columnsep=15pt]{geometry} 
\usepackage{adjustbox}
\usepackage{placeins}
\usepackage[T1]{fontenc}
\usepackage{lipsum}
\usepackage{csquotes}
\usepackage[pdftex, pdftitle={Article}, pdfauthor={Author}]{hyperref} % For hyperlinks in the PDF
\usepackage{bm}% bold math
\providecommand{\abs}[1]{$\left|#1\right|$}

\providecommand{\ket}[1]{|#1\rangle}
\providecommand{\bra}[1]{\langle#1|}
\providecommand{\brak}[2]{\langle#1|#2\rangle} % si usa come: \brak{A}{B} = <A|B>
\providecommand{\proj}[2]{|#1\rangle \! \langle#2|} % si usa come: \proj{A}{B} = |A><B|
\providecommand{\mean}[3]{\langle#1|#2|#3\rangle} % si usa come: \mean{A}{B}{C} = <A|B|C>
\newcommand{\intR}{{\int_{\mathbb{R}^2}}}

\begin{document}
%\title{Splitting phenomenon in non--separable beams}
\title{Measuring the non-separability of spatially disjoint vectorial fields}
\author{Andrea Aiello}
\affiliation{Max Planck Institute for the Science of Light, Staudtstrasse 2, 91058 Erlangen, Germany}
\author{Xiao-Bo Hu}
\affiliation{Key Laboratory of Optical Field Manipulation of Zhejiang Province,Department of Physics, Zhejiang Sci-Tech University, Hangzhou, 310018, China.}
\author{Valeria Rodr\'iguez-Fajardo}
\affiliation{School of Physics, University of the Witwatersrand, Private Bag 3, Wits 2050, South Africa}
\author{Raul I. Hernandez-Aranda}
\affiliation{Photonics and Mathematical Optics Group, Tecnologico de Monterrey, Monterrey 64849, Mexico.}
\author{Andrew Forbes}
\affiliation{School of Physics, University of the Witwatersrand, Private Bag 3, Wits 2050, South Africa}
\author{Benjamin Perez-Garcia}
\email[Correspondence email address: ]{b.pegar@tec.mx}
\affiliation{Photonics and Mathematical Optics Group, Tecnologico de Monterrey, Monterrey 64849, Mexico.}

\author{Carmelo Rosales-Guzm\'an}
\email[Correspondence email address: ]{carmelorosalesg@cio.mx}% Your name
\affiliation{Centro de Investigaciones en \'Optica, A.C., Loma del Bosque 115, Colonia Lomas del campestre, C.P. 37150 Le\'on, Guanajuato, Mexico}
\affiliation{Wang Da-Heng Collaborative Innovation Center, Heilongjiang Provincial Key Laboratory of Quantum Manipulation and Control, Harbin University of Science and Technology, Harbin 150080, China.}

\date{\today} % Leave empty to omit a date

%%%%%%%%%%%%%%%%%%%%%%%%%%%%%%%%%%%%%%%%
\begin{abstract}
%%%%%%%%%%%%%%%%%%%%%%%%%%%%%%%%%%%%%%%%
\noindent Vectorial forms of structured light that are non-separable in their spatial and polarisation degrees of freedom have become topical of late, with an extensive toolkit for their creation and control. In contrast, the toolkit for quantifying their non-separability, the inhomogeneity of the polarisation structure, is far less developed, and in some cases fails altogether.  To overcome this, here we introduce a new measure for vectorial light, which we demonstrate both theoretically and experimentally.  We consider the general case where the local polarisation homogeneity can vary spatially across the field, from scalar to vector, a condition that can arise naturally if the composite scalar fields are path separable during propagation, leading to spatially disjoint vectorial light.  We show how the new measure correctly accounts for the local path-like separability of the individual scalar beams, which can have varying degrees of disjointness, even though the global vectorial field remains intact. Our work attempts to address a pressing issue in the analysis of such complex light fields, and raises important questions on spatial coherence in the context of vectorially polarised light. 
\end{abstract}


\maketitle

%%%%%%%%%%%%%%%%%%%%%%%%%%%%%%%%%%%%%%%%
\section{Introduction}
%%%%%%%%%%%%%%%%%%%%%%%%%%%%%%%%%%%%%%%%
Complex vectorial fields are general states of structured light \cite{Forbes2021}, non-separable in their spatial and polarisation degrees of freedom, giving rise to exotic inhomogeneous transverse polarisation distributions \cite{Galvez2012}. These vectorial forms of structured light have shown their potential in a wide variety of applications \cite{Ndagano2018IEEE,Yang2021,Hu2019,BergJohansen2015,Toppel2014,Rosales2018Review}, including optical communications, optical tweezers, optical metrology, amongst others. Noteworthy, their high resemblance to quantum-entangled states by virtue of the non-separability of their component degrees of freedom has been the subject of several studies, which have enabled simulating quantum phenomena in the classical regime \cite{Spreeuw1998,Aiello2015,Ndagano2017,Guzman-Silva2016,Balthazar2016,forbes2019classically,konrad2019}.
%Naidoo2016,sroor2020high,
Given the high interest in vector beams, in the last two decades there has been intense interest in their generation, with techniques exploiting liquid crystal wave plates \cite{Marrucci2006}, glass cones \cite{Radwell2016,Kozawa2005}, metamaterials \cite{Devlin2017}, interferometric arrays \cite{Tidwell1990,Niziev2006,Passilly2005,Mendoza-Hernandez2019}, directly from lasers \cite{forbes2019structured}, liquid crystal spatial light modulators  \cite{Davis2000,Maurer2007,Moreno2012,Mitchell2017,SPIEbook,Rosales2017,Rong2014,Liu2018} and digital micromirror devices 
\cite{Ren2015,Mitchell2016,Scholes2019,Gong2014,Rosales2020,Rosales2021,XiaoboHu2021,Yao-Li2020}. Conversely, characterisation techniques are still somewhat limited.  These include filtering \cite{ndagano2015fiber,milione2015using,milione20154} and deterministic detection \cite{ndagano2017deterministic} of the mode, as well as quantitative analysis by a quantum toolkit, for a full decomposition via tomographic projections \cite{toninelli2019concepts} and a reduced measurement set for determining the non-separability \cite{McLaren2015,Ndagano2016,Zhaobo2020,Selyem2019}.  Both quantitative techniques rely on the similarities between the vector modes and quantum-entangled states, for example, the latter being the classical equivalent to the well-known concurrence, $C$, for qubits \cite{Wootters1998,Wootters2001}.  In the classical case, rather than measuring the degree of entanglement of two entangled photons, the measurement has been adapted to measure the degree of coupling between the spatial and polarisation degrees of freedom of vector modes, and has been coined the Vector Quality Factor (VQF) \cite{Ndagano2016}. 

\begin{figure}[tb]
    \centering
    \includegraphics[width=0.4\textwidth]{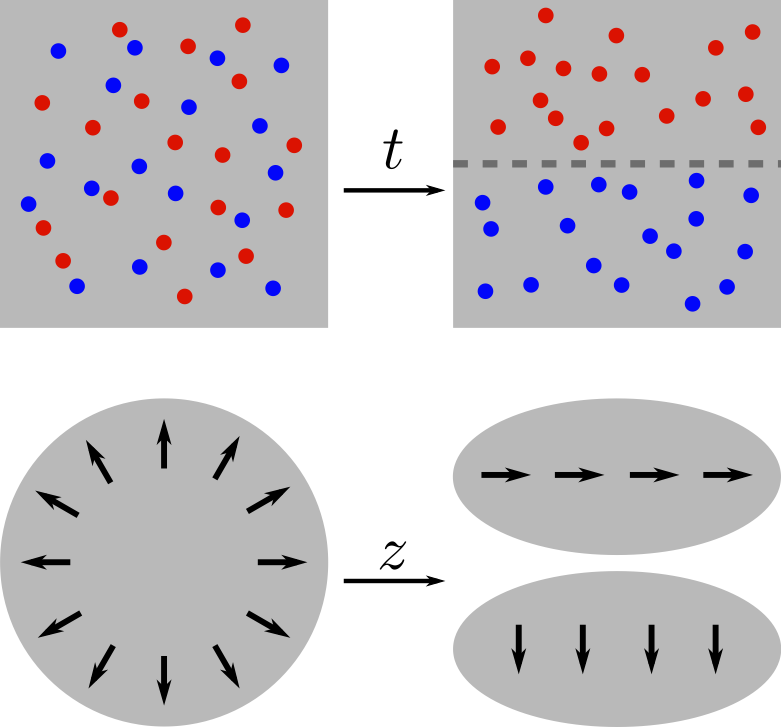}
    \caption{Conceptual figure.  The upper row shows a mixture of two different kind of particles (red and blue types).  As time evolves, each kind of particle groups together in separate regions.  The lower row depicts a vectorial field that upon propagation splits into disjoint areas.}
    \label{fig:concept}
\end{figure}

Even though these techniques perform very well at characterising some features of vector beams, such as their non-separability, they fail to account for other properties manifested by more intricate vector modes.  One such case is when the spatial modes become spatially disjoint, appearing to follow independent paths although remaining as a single coherent field.  This situation can arise naturally in vectorial light and has been discussed theoretically \cite{Aiello2015} and observed experimentally \cite{XiaoboHu2021}, and even engineered as a path degree of freedom in such fields \cite{shen2020structured,shen2021creation,pabon2020high}. This path-like splitting is analogous to splitting of particle mixtures, illustrated in Fig.\,\ref{fig:concept}. In the case of such splitting phenomena in vectorial light, the mutable degree of non-separability cannot be quantified using conventional measurements such as the VQF, suggesting the need for a more refined toolbox. 

Here we introduce a novel approach for the characterisation of vectorial light fields that overcomes the aforementioned shortcoming.  We outline the problem theoretically and propose an amendment to the existing framework to account not only for the non-separability of the entire field, but also its component disjointness.  For this, we employ the Hellinger distance \cite{Hellinger} as a measure of the disjointness.  We provide experimental evidence to verify the efficacy of this technique, using exemplar cases of Hermite- Airy-, and Parabolic-Gauss vectorial beams.

%%%%%%%%%%%%%%%%%%%%%%%%%%%%%%%%%%%%%%%%
\section{Theory}\label{t1}
%%%%%%%%%%%%%%%%%%%%%%%%%%%%%%%%%%%%%%%%
In this section, we provide the theoretical background for describing and quantifying the splitting of the polarisation pattern of a non-uniformly polarised light beam into two disjoint uniformly polarised parts, which may occur during free propagation, as sketched conceptually in Fig.\,\ref{fig:concept}. After shortly reviewing in Sec. \ref{cursory} the basic properties of non-uniformly polarised beams of light, we give in Sec. \ref{dop} a description of the above splitting phenomenon in terms of the degree of polarisation of each part of the beam. Then, in Sec. \ref{hellingerdistance} we introduce a propagation-dependent measure of this phenomenon, which is suitable for experimental verification. Finally, we illustrate our approach by means of a real-world example in Sec. \ref{example}.

%%%%%%%%%%%%%%%%%%%%%%%%%%%%%%%%%%%%%%%%
\subsection{Entangled polarisation and spatial degrees of freedom in optical beams: a cursory look}\label{cursory}

Consider a monochromatic, non-uniformly polarised paraxial beam of light travelling in vacuum (or  air) along the positive direction of the  axis $z$ of a Cartesian reference frame with coordinates $\bm{r} = (x,y,z)= (\bm{x},z)$, where $\bm{x} = (x,y)$. The measurable real-valued electric field vector $\mathbf{E}(\bm{r},t)$ carried by such beam can be written as
\begin{align}\label{a10}
\mathbf{E}(\bm{r},t) = \mathrm{Re} \left[ \bm{\Psi}(\bm{x},z)\, e^{i(k z - \omega t)} \right] ,
\end{align}
where $k = 2 \pi /\lambda$ is the wavenumber of the light of wavelength $\lambda$ and frequency $\omega = k c$, with $c$  the speed of light in vacuum.
By definition, the vector  field $\bm{\Psi}(\bm{x},z)$ can always be written as
\begin{align}\label{a20}
\bm{\Psi}(\bm{x},z) = \hat{\bm{e}}_1 \, \psi_1(\bm{x},z) + \hat{\bm{e}}_2 \, \psi_2(\bm{x},z),
\end{align}
where  $\hat{\bm{e}}_1,\hat{\bm{e}}_2$ and $\hat{\bm{e}}_3$ are  real-valued unit vectors parallel to the axes  $x,y$ and  $z$, respectively,  with
\begin{align}\label{a30}
\hat{\bm{e}}_i \cdot  \hat{\bm{e}}_j = \delta_{ij}, \qquad (i,j=1,2,3),
\end{align}
and $\psi_1(\bm{x},z),\psi_2(\bm{x},z)$ are two arbitrary solutions of the paraxial wave equation,
\begin{align}\label{a40}
\left( \frac{\partial^2}{\partial x^2} + \frac{\partial^2}{\partial y^2} + 2 i k \frac{\partial}{\partial z} \right)\psi_\alpha(\bm{x},z)=0, \quad (\alpha = 1,2).
\end{align}
Alternatively, as shown in Appendix \ref{appSchmidt}, it is possible to write $\bm{\Psi}(\bm{x},z)$ in the Schmidt  form \cite{Aiello2015}, %\ref{Schmidt}),
\begin{align}\label{a42}
\bm{\Psi}(\bm{x},z) = \sqrt{\lambda_1} \, \hat{\bm{\epsilon}}_1 \, \phi_1(\bm{x},z) + \sqrt{\lambda_2} \, \hat{\bm{\epsilon}}_2 \, \phi_2(\bm{x},z),
\end{align}
where $\lambda_1 \geq \lambda_2 \geq 0$, are the eigenvalues of the coherency matrix $J$ of the beam (see Eq. \eqref{similar} in Appendix \ref{tre}), defined by
\begin{align}\label{a44}
J =  \int_{\mathbb{R}^2} \mathrm{d}^2 x \, \bm{\Psi}\bm{\Psi}^\dagger.
\end{align}
The dyad $\bm{\Psi}\bm{\Psi}^\dagger$ is an $\bm{x}$- and $z$-dependent $2 \times 2$ matrix which can be written with respect to the basis $\{\hat{\bm{e}}_1, \hat{\bm{e}}_2 \}$ as:
\begin{align}\label{a45}
\bm{\Psi}\bm{\Psi}^\dagger =
\begin{bmatrix}
|\psi_1(\bm{x},z)|^2 & \psi_1(\bm{x},z) \psi_2^*(\bm{x},z) \\[6pt]
\psi_1^*(\bm{x},z) \psi_2(\bm{x},z) & |\psi_2(\bm{x},z)|^2 \\
\end{bmatrix},
\end{align}
where Eq. \eqref{a20} has been used. We remark that $J$ is independent of $z$ because both $\psi_1(\bm{x},z)$ and $\psi_2(\bm{x},z)$ are supposedly solutions of the paraxial wave equation.
 The polarisation modes $\hat{\bm{\epsilon}}_1$ and $\hat{\bm{\epsilon}}_2$  are two (possibly complex-valued) unit vectors perpendicular to the $z$-axis and to each other,
\begin{align}\label{a50}
 \hat{\bm{\epsilon}}_\alpha^* \cdot  \hat{\bm{e}}_3   =0, \quad \hat{\bm{\epsilon}}_\alpha^* \cdot  \hat{\bm{\epsilon}}_\beta   = \delta_{\alpha \beta},
\end{align}
with $\alpha,\beta = 1,2$.
Likewise, the spatial modes $\phi_1(\bm{x},z)$ and  $\phi_2(\bm{x},z)$ are square-integrable solutions of the paraxial wave equation  that  obey the orthonormality conditions
\begin{align}\label{a60}
\int_{\mathbb{R}^2} \mathrm{d}^2 x \,  \phi_\alpha^*(\bm{x},z)  \phi_\beta(\bm{x},z)  =  \delta_{\alpha \beta},  \quad (\alpha,\beta = 1,2).
\end{align}

The  entanglement between the polarisation and the spatial degrees of freedom of the vector field \eqref{a20} is usually quantified by the concurrence $C$ defined by
\begin{align}\label{a70}
C = 2 \, \sqrt{\vphantom{I}p_1  \, p_2} , \qquad \text{where} \qquad p_\alpha \equiv \frac{\lambda_\alpha}{\lambda_1 + \lambda_2},
\end{align}
and $\alpha = 1,2$. By definition,
the concurrence is related to the degree of polarisation $P$ of the beam \cite{BW,Neill,PhysRevA.89.013845},
\begin{align}\label{a72}
P = \sqrt{1 - \frac{4 \det J}{\left(\operatorname{tr} J \right)^2}}  = |p_1- p_2|,
\end{align}
by the simple relation
\begin{align}\label{a74}
C^2 + P^2 = 1.
\end{align}
When either $p_1=0$ or $p_2=0$, there is not entanglement so that $C=0$ and $P=1$. Vice versa, when $p_1 = p_2 = 1/2$ the entanglement is maximal and we obtain $C=1$ and $P=0$.
From the definition \eqref{a45} of $J$, it follows that the concurrence $C$ is propagation-invariant,  i.e., it does not depend on the propagation distance $z$, even if the polarisation pattern changes upon propagation.

%%%%%%%%%%%%%%%%%%%%%%%%%%%%%%%%%%%%%%%%
\subsection{Polarisation-pattern splitting and degree of polarisation}\label{dop}

When a non-uniformly polarised paraxial beam of light travels in free space, its polarisation pattern typically changes during propagation, although the concurrence and the degree of polarisation do remain constant (see, e.g., \cite{photonics8110491} and references therein). Of particular interest are those peculiar beams whose polarisation pattern splits upon propagation  into two disjoint parts with uniform orthogonal polarisation, as outlined in Fig.\,\ref{fig:concept} \cite{XiaoboHu2021}. Let $z = z_0$ be the distance from the origin of the coordinates at which the split occurs, and let $\hat{\bm{\epsilon}}_1$ and $\hat{\bm{\epsilon}}_2$ denote the unit vectors characterising the orthogonal polarisations of the two parts of the beam resulting from the splitting. Note that $z_0$ may be either finite or infinite as, for example, in the case of the beams investigated in \cite{XiaoboHu2021}. Then, the field of the beam at $z=z_0$ must have necessarily the form
\begin{align}\label{dop10}
\bm{\Psi}(\bm{x},z_0) =  \hat{\bm{\epsilon}}_1 \, \varphi_1(\bm{x},z_0) +  \hat{\bm{\epsilon}}_2 \, \varphi_2(\bm{x},z_0),
\end{align}
with
\begin{align}\label{dop20}
\varphi_1(\bm{x},z_0) \, \varphi_2(\bm{x},z_0) = 0, \qquad \forall \bm{x} \in \mathbb{R}^2.
\end{align}
The last condition is implied by the requirement that at $z = z_0$ the two  polarisation patterns do not overlap at any point in the $xy$-plane. In more mathematical terms, Eq. \eqref{dop20} can be written as
\begin{align}\label{dop22}
\mathcal{D}_1 \cap \mathcal{D}_2 = \emptyset,
\end{align}
where $\mathcal{D}_\alpha $ denotes the support of the mode function $\varphi_\alpha$ at $z = z_0$, namely
\begin{align}\label{dop60}
\mathcal{D}_\alpha = \operatorname{supp} \left( \varphi_\alpha \right) =  \left\{ \bm{x} \in \mathbb{R}^2 : \varphi_\alpha( \bm{x}, z_0 ) \neq 0 \right\}.
\end{align}
Here and hereafter we assume that $\mathcal{D}_1 \cup \mathcal{D}_2 = \mathbb{R}^2$.

From \eqref{dop20} it trivially follows that the spatial modes $\varphi_1(\bm{x},z_0)$ and $\varphi_2(\bm{x},z_0)$ are orthogonal in the sense of \eqref{a60}. Therefore, Eq. \eqref{dop10} is automatically a Schmidt  form  of the type \eqref{a42}, with
\begin{align}\label{dop30}
\sqrt{\lambda_\alpha} \, \phi_\alpha(\bm{x},z_0) = \varphi_\alpha(\bm{x},z_0),
\end{align}
and
\begin{align}\label{dop40}
\lambda_\alpha =  \int_{\mathbb{R}^2} \mathrm{d}^2 x \,\left| \varphi_\alpha(\bm{x},z_0) \right|^2 .
\end{align}
Since  paraxial propagation is a unitary process, orthogonality is preserved during propagation, so that
\begin{align}\label{dop46}
\int_{\mathbb{R}^2} \mathrm{d}^2 x \,  \varphi_\alpha^*(\bm{x},z) \varphi_\beta(\bm{x},z)  = \lambda_\alpha \, \delta_{\alpha \beta},  \quad (\alpha,\beta = 1,2),
\end{align}
where \eqref{dop40} has been used, and
\begin{align}\label{dop42}
\varphi_\alpha(\bm{x},z) = \int_{\mathbb{R}^2} \mathrm{d}^2 x' \,  U \left( \bm{x}- \bm{x}', z - z_0 \right) \varphi_\alpha(\bm{x}',z_0),
\end{align}
with
\begin{align}\label{dop44}
U \left( \bm{x}- \bm{x}', z \right) = \frac{k}{2 \pi  i} \frac{1}{z} \, \exp\left[i \, \frac{k}{2 z} \left(\bm{x} - \bm{x}' \right)^2 \right],
\end{align}
the Fresnel paraxial propagator \cite{Goodman2017}.
This implies that also the Schmidt decomposition \eqref{dop10} remains valid, with
\begin{align}\label{dop10bis}
\bm{\Psi}(\bm{x},z) =  \hat{\bm{\epsilon}}_1 \, \varphi_1(\bm{x},z) +  \hat{\bm{\epsilon}}_2 \, \varphi_2(\bm{x},z).
\end{align}

Another useful decomposition of $\bm{\Psi}(\bm{x},z)$  is the following,
\begin{align}\label{dop62}
\bm{\Psi}(\bm{x},z) =  \bm{\Psi}_1(\bm{x},z) + \bm{\Psi}_2(\bm{x},z),
\end{align}
where
\begin{align}\label{dop90}
\bm{\Psi}_\mu(\bm{x},z) \equiv &  \left. \vphantom{\Big[} \bm{\Psi}(\bm{x},z) \right|_{\bm{x} \in \mathcal{D}_\mu} \nonumber \\[4pt]
= & \, \hat{\bm{\epsilon}}_1  \left. \vphantom{\Big[} \varphi_1(\bm{x},z)\right|_{\bm{x} \in \mathcal{D}_\mu}  +  \hat{\bm{\epsilon}}_2  \left. \vphantom{\Big[} \varphi_2(\bm{x},z)\right|_{\bm{x} \in \mathcal{D}_\mu},
\end{align}
with $\mu=1,2$, denotes the field $\bm{\Psi}(\bm{x},z)$ restricted to the region $\mathcal{D}_\mu$ of the $xy$-plane. At $z = z_0$, Eq. \eqref{dop90} reduces to
\begin{align}\label{dop90bis}
\bm{\Psi}_\mu(\bm{x},z_0) =  \hat{\bm{\epsilon}}_\mu  \varphi_\mu (\bm{x},z_0), \qquad (\mu=1,2).
\end{align}

Both fields $\bm{\Psi}_1(\bm{x},z_0)$ and $\bm{\Psi}_2(\bm{x},z_0)$ defined by \eqref{dop90bis}, are fully polarised so that their degree of polarisation  is equal to one.
However, for $z \neq z_0$,  $\bm{\Psi}_1(\bm{x},z)$ and $\bm{\Psi}_2(\bm{x},z)$
are typically partially polarised  because of the physical superposition between $\left.\varphi_1(\bm{x},z)\right|_{\bm{x} \in \mathcal{D}_\mu}$ and $ \left.\varphi_2(\bm{x},z)\right|_{\bm{x} \in \mathcal{D}_\mu}$, which may occur because of diffraction. In this case, the degree of polarisation of each field must be less than one. This implies that the degree of polarisation of $\bm{\Psi}_\mu(\bm{x},z)$, denoted $P_\mu(z)$, will depend on the propagation distance $z$ and can therefore be used as a measure of the polarisation-pattern splitting phenomenon.
$P_\mu(z)$ can be calculated either via \eqref{a72}, or as $P = \sqrt{1-C^2}$, where  \eqref{a74} has been used. Then, using the expression \eqref{l170} for $C$, we can eventually write
\begin{align}\label{dop110}
P_\mu(z) =  \frac{\sqrt{  S_{1 \mu}^2(z)  + S_{2 \mu}^2(z) + S_{3 \mu}^2(z)}}{S_{0 \mu}(z)} \,  ,
\end{align}
where  the measurable Stokes parameters of the field $\bm{\Psi}_\mu(\bm{x},z)$, are given by
\begin{equation}\label{dop105}
\begin{split}
S_{0 \mu}(z) = & \;  \int_{\mathcal{D}_\mu} \mathrm{d}^2 x \left( \left| \psi_1 \right|^2 + \left| \psi_2 \right|^2 \right), \\[4pt]
S_{1 \mu}(z) = & \;  \int_{\mathcal{D}_\mu} \mathrm{d}^2 x \left( \left| \psi_1 \right|^2 - \left| \psi_2 \right|^2 \right),  \\[4pt]
S_{2 \mu}(z) = & \;  \int_{\mathcal{D}_\mu} \mathrm{d}^2 x \left( \psi_1^* \psi_2 + \psi_1 \psi_2^* \right) , \\[4pt]
S_{3 \mu}(z) = & \;  \frac{1}{i}\int_{\mathcal{D}_\mu} \mathrm{d}^2 x \left( \psi_1^* \psi_2 - \psi_1 \psi_2^* \right),
\end{split}
\end{equation}
with $\psi_\alpha = \psi_\alpha(\bm{x},z) =  \hat{\bm{e}}_\alpha \cdot \bm{\Psi}(\bm{x},z)$, and $\alpha = 1,2$.

By construction, at $z=z_0$ we must have $P_1(z_0)= 1 = P_2(z_0)$. However, for $z \neq z_0$ we expect to find $P_\mu(z) < 1$. An explicit example of application of Eq. \eqref{dop110} will be presented in Sec. \ref{example}.

%%%%%%%%%%%%%%%%%%%%%%%%%%%%%%%%%%%%%%%%
\subsection{The Hellinger distance}\label{hellingerdistance}

Let us look again at the  field \eqref{dop10bis}, that we rewrite here as
\begin{align}\label{hd10}
\bm{\Psi}(\bm{x},z) =  \sqrt{\lambda_1} \, \hat{\bm{\epsilon}}_1 \, \phi_1(\bm{x},z) +  \sqrt{\lambda_2} \, \hat{\bm{\epsilon}}_2 \, \phi_2(\bm{x},z),
\end{align}
where, according to Eqs. \eqref{dop30} and \eqref{dop42}, we have defined
\begin{align}\label{hd20}
 \phi_\alpha(\bm{x},z) = \frac{\varphi_\alpha(\bm{x},z)}{\sqrt{\lambda_\alpha} }, \qquad (\alpha=1,2),
\end{align}
with
\begin{align}\label{hd30}
\phi_\alpha(\bm{x},z_0) \phi_\beta^*(\bm{x},z_0) = \delta_{\alpha \beta} |\phi_\alpha(\bm{x},z_0)|^2.
\end{align}
Now, consider the non-negative quantity
\begin{align}\label{hd40}
h[\phi_1,\phi_2](z) = \int_{\mathbb{R}^2} \text{d}^2 x \, \left| \phi_1(\bm{x},z) \right|  \left| \phi_2(\bm{x},z) \right| \geq 0 ,
\end{align}
which is a functional of $\phi_1$ and $\phi_2$, and a function of $z$.
The minimum value
\begin{align}\label{hd50}
\min \bigl\{ h[\phi_1,\phi_2](z) \bigr\} = h[\phi_1,\phi_2](z_0) = 0,
\end{align}
is achieved at $z=z_0$ because of Eq. \eqref{hd30}. The maximum value
\begin{align}\label{hd60}
\max \bigl\{  h[\phi_1,\phi_2](z) \bigr\} = h[\phi_1,\phi_1](z)= 1,
\end{align}
is obtained for $\left| \phi_1(\bm{x},z) \right| = \left| \phi_2(\bm{x},z) \right|$, because of the normalization condition \eqref{dop46}.
Next, we note that the non-negative function $f_\alpha(\bm{x},z)$, defined by
\begin{align}\label{hd70}
f_\alpha (\bm{x},z) \equiv   \left| \phi_\alpha(\bm{x},z) \right|^2 \geq 0, \qquad (\alpha =1,2),
\end{align}
has all the properties of the probability density function of a continuous real-valued two-dimensional random vector $X = (X_1,X_2)$, with $(X_1,X_2) \sim f_\alpha (x_1,x_2,z)$.
In probability theory and statistics, there are several ways to quantify the distance between two probability distributions, say $f(x)$ and $g(x)$. Here, considering the properties (\ref{hd50}-\ref{hd60}), it is convenient to use the  Hellinger distance $H[f,g]$ \cite{Hellinger}, defined by
\begin{align}\label{hd80}
H^2[f,g] = & \; \frac{1}{2}\int \mathrm{d} x \, \left[ \sqrt{f(x)} - \sqrt{g(x)}\right]^2   \nonumber \\[6pt]
= & \; 1 - \int \mathrm{d} x \, \sqrt{f(x) g(x)} \, .
\end{align}
It is not difficult to show that such distance is a proper \emph{metric}, that is  for all non-negative, normalized  smooth functions $g_1, g_2 , g_3$, the Hellinger distance $H[g_1,g_2]$ satisfies the following conditions:
\begin{enumerate}
\setlength\itemsep{-0.1truecm}
  \item positivity condition: $H[g_1,g_2] \geq 0$ ;
  \item symmetry property: $H[g_1,g_2] = H[g_2,g_1]$;
  \item identity property: $H[g_1,g_2] = 0$ if and only if $g_1=g_2$;
  \item triangular inequality: $H[g_1,g_3] \leq  H[g_1,g_2] + H[g_2,g_3]$.
\end{enumerate}

In our two-dimensional case the Hellinger distance takes the form
\begin{align}\label{hd90}
H[|\phi_1|,|\phi_2|](z) = & \;  \sqrt{1 -   \int_{\mathbb{R}^2} \left| \phi_1(\bm{x},z) \right|  \left| \phi_2(\bm{x},z) \right| \, \text{d}^2 x}  \nonumber \\[8pt]
= & \;  \sqrt{ \vphantom{\Big[} 1 -   h[\phi_1,\phi_2](z)} \,.
\end{align}
We remark that this distance does not depend directly on the amount of the entanglement of the beam, that it is independent of $\lambda_1$ and $\lambda_2$. However, it does crucially depend on the propagation distance $z$.  For illustration,  $H[|\phi_1|,|\phi_2|](z)$ is calculated explicitly for an exemplary beam in Sec. \ref{example}.

%%%%%%%%%%%%%%%%%%%%%%%%%%%%%%%%%%%%%%%%
\subsection{Polarisation-pattern splitting phenomenon: A case study}\label{example}

To illustrate the use of Eqs. \eqref{dop110} and \eqref{hd90}, we consider here a simple non-uniformly polarised paraxial beam of light, whose polarisation pattern is split into two disjoint uniformly polarised parts at $z= z_0 =0$.
Imagine to have a beam of light prepared at $z=0$ in the Hermite-Gauss mode $u_{01}(\bm{x},0)$ of Rayleigh length $z_R$ and waist $w_0$ (see Appendix \ref{HGmodes}). Then, suppose to place at $z=0$ a thin spherical lens of focal length $f= z_R$,  characterised by the transmission function \cite{Goodman2017},
\begin{align}\label{a270}
t(\bm{x}) = e^{- i k \, \frac{\left|\bm{x} \right|^2}{2f} } = e^{- i  \, \frac{\left|\bm{x} \right|^2}{w_0^2} }.
\end{align}
By construction, the amplitude $\phi(\bm{x})$ of the beam at $z=0$ immediately after the lens is
\begin{align}\label{a280}
\phi(\bm{x}) = t(\bm{x}) \, u_{01}(\bm{x},0) = \sqrt{\frac{8}{\pi}} \, \frac{y}{w_0^2} \, e^{-(1+ i)  \, \frac{\left|\bm{x} \right|^2}{w_0^2} }.
\end{align}
Using $\phi(\bm{x})$, we can define the two orthonormal mode functions $\phi_1$ and $\phi_2$ at $z=0$, as
\begin{subequations}\label{a290}
\begin{align}
\phi_1(\bm{x},0) = & \; \sqrt{2} \, \phi(\bm{x}) \, \theta(y) , \label{a290A} \\[6pt]
 \phi_2(\bm{x},0) = & \; \sqrt{2} \, \phi(\bm{x})\,  \theta(-y), \label{a290B}
\end{align}
\end{subequations}
where $\theta(y)$ denotes the Heaviside step function \cite{Hfunction}. Finally, imagine to send the upper (lower) mode $\phi_1$ ($\phi_2$) through a linear polarizer  and a quarter wave plate suitably oriented, so that it is circularly polarised to the right (left). In this way, we obtain a non-uniformly polarised beam with vector amplitude %of which can be written  as
\begin{align}\label{a300}
\bm{\Psi}(\bm{x},0) = \phi(\bm{x}) \bigl[ \hat{\bm{\epsilon}}_1 \,  \theta(y) + \hat{\bm{\epsilon}}_2 \, \theta(-y) \bigr],
\end{align}
where the unit vectors
\begin{align}\label{a310}
\hat{\bm{\epsilon}}_\alpha = \frac{\hat{\bm{e}}_1 + i (-1)^\alpha \hat{\bm{e}}_2}{\sqrt{2}}, \qquad (\alpha=1,2),
\end{align}
describe right- ($\alpha = 1$) and left-circular ($\alpha = 2$) polarisation. By definition, this field is maximally entangled and $C=1$.
The vector field $\bm{\Psi}(\bm{x},z)$ at distance $z$ from the origin can be calculated as in Eq. \eqref{dop10bis}.

\begin{figure}[htbp]
  \centering
 \includegraphics[scale=3,clip=false,width=0.8\columnwidth,trim = 0 0 0 0]{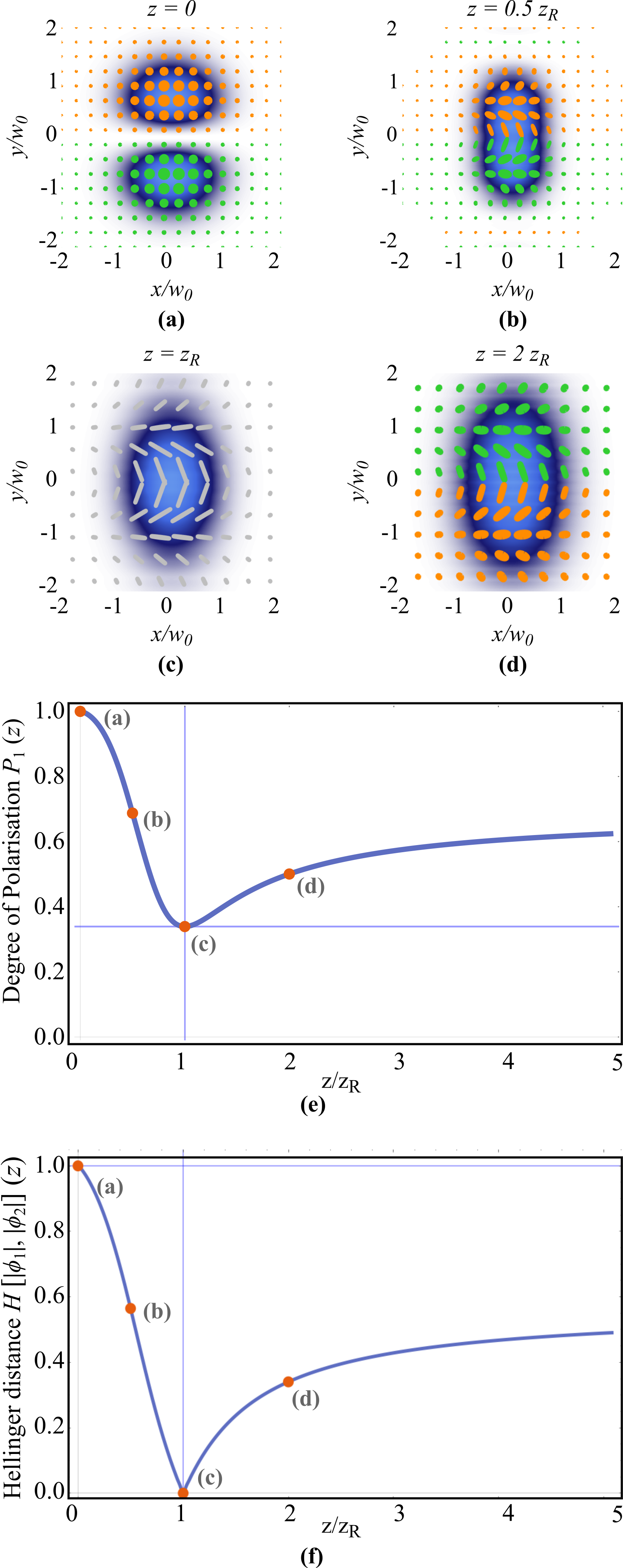}
  \caption{Polarisation-intensity patterns of the mode defined by Eq. \eqref{a300} at (a) $z=0$, (b) $z=0.5 z_R$, (c) $z=z_R$ and (d) $z=2z_R$. Right and left circular polarisation are shown in orange and green, respectively. Notice the top-bottom inversion of handedness occurring from $z< z_R$ to $z>z_R$. (e) and (f) show the degree of polarisation $P_1(z)$ and Hellinger distance $H[|\phi_1|, |\phi_2|](z) $, respectively, as function of propagation distance. Note that $H[|\phi_1|, |\phi_2|](z)$ is maximum when the two spatial modes are non-overlapping ($z=0$) and minimum when their overlap is maximal ($z = z_R$), which also coincides with a totally scrambled polarisation distribution. The red dots labelled \textsf{(a,b,c,d)} mark the distances $z/z_R = 0, 0.5, 1,2$, respectively, corresponding to the polarisation patterns shown.}\label{hellinger}
\end{figure}

% The minimum value  $P_1(z=z_R) \cong 0.338$ is attained at $z=z_R$. The most significant part of the plot is located at $0 \leq z \leq z_R$, where we pass from two fully disjoint spatial modes at $z=0$, to two maximally overlapping modes at $z=z_R$.

The polarisation and intensity patterns of $\bm{\Psi}(\bm{x},z)$ at several distances $z \geq 0$ are shown in Fig.\,\ref{hellinger} (a)-(d). At $z=0$ (see Fig.\,\ref{hellinger} \textsf{(a)}) the beam presents two disjoint polarisation patterns. Upon propagation by $z$, the upper and lower lobes of the field  \eqref{a310} spread and overlap, and at $z = z_R$ the  polarisation pattern becomes completely scrambled, as shown in  Fig.\,\ref{hellinger} (c).

In Fig.\,\ref{hellinger} (e) we show the degree of polarisation $P_1(z)$ of the beam $\bm{\Psi}_1(\bm{x},z) = \left. \bm{\Psi}(\bm{x},z) \right|_{\bm{x} \in \mathcal{D}_1}$, where $\mathcal{D}_1$ coincides with the upper half plane $y\geq 0$. For this beam $P_1(z)=P_2(z)$. As expected, at $z = z_0=0$ the fields $\bm{\Psi}_1(\bm{x},0)$ and $\bm{\Psi}_2(\bm{x},0)$ are fully polarized. Then, the degree of polarisation decrease with $z$ to reach the minimum value at $z = z_R$, where $\left.\varphi_1(\bm{x},z)\right|_{\bm{x} \in \mathcal{D}_\mu}$ and $ \left.\varphi_2(\bm{x},z)\right|_{\bm{x} \in \mathcal{D}_\mu}$ are maximally overlapped. Finally, for $z>z_R$ the fields tend to separate again and $P_1(z)$ raises monotonically.

Figure\,\ref{hellinger} (f) shows the plot of the Hellinger distance $H[|\phi_1|,|\phi_2|](z)$ calculated from Eq. \eqref{hd90}. It is worth noticing that the plots of $H[|\phi_1|,|\phi_2|](z)$  and  $P_1(z)$ display the same qualitative behaviour. The Hellinger distance is maximal  when the two polarisation patterns are fully separated. Vice versa, $H[|\phi_1|,|\phi_2|](z)$ is minimal at $z = z_R$ where the patterns are maximally superimposed.

\section{Experiment}\label{Exp}  % Carmelo, Valeria
%%%%%%%%%%%%%%%%%%%%%%%%%%%%%%%%%%%%%%%%

%%%%%%%%%%%%%%%%%%%%%%%%%%%%%%%%%%%%%%%%
\subsection{Experimental setup}

\begin{figure}[tb]
    \centering
    \includegraphics[width=0.48\textwidth]{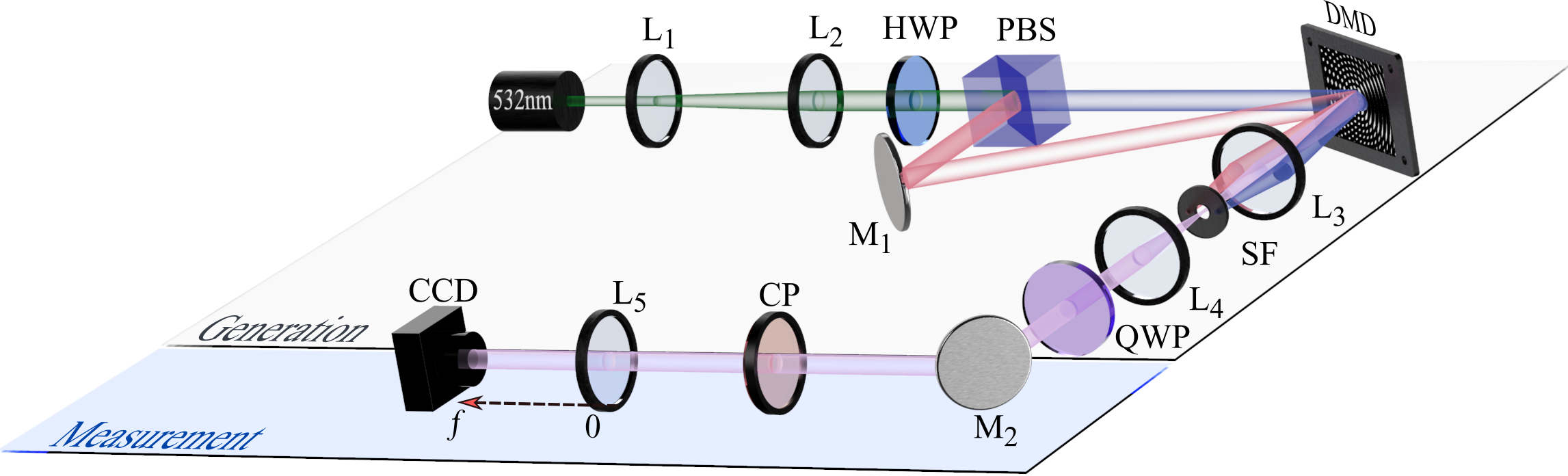}
    \caption{Experimental setup to generate and characterise no-separable light beams. In the {\bf Generation} stage, an expanded and collimated diagonally polarised laser beam ($\lambda=532$ nm) is split into its two polarisation components using a Polarising Beam Splitter (PBS). Both beams are redirected with the help of a mirror to the centre of a Digital Micromirror Device (DMD), where the require binary holograms are displayed to generate the desired non-separable light modes. L: Lens; HWP and QWP: Quarter- and Half-Wave Plate, respectively; M: Mirror; SF: Spatial Filter; CP: Circular Polariser. In the {\bf Measurement} stage, the modes are analysed using a Charge-Coupled Device (CCD) camera mounted on a translation stage. Further, the generated modes are focused with a lens (L$_5$) of focal distance $f=400$ mm.}
    \label{setup}
\end{figure}

To experimentally prove our proposed theory, we implemented the experimental setup previously demonstrated in \cite{Rosales2020} and schematically depicted in Fig.\,\ref{setup} in the {\bf Generation} stage. Here, a horizontally polarised laser beam ($\lambda=532$ nm, 300 mW) is expanded and collimated with the telescope formed by the pair of lenses $\rm L_1$ ($f_1=20$ mm) and $\rm L_2$  ($f_2=200$ mm). The polarisation state of the beam is then rotated to a diagonal polarisation state through a Half-Wave Plate (HWP) oriented at $22.5^\circ$ relative to the vertical axis. Afterwards, the beam is split, using a Polarising Beam Splitter (PBS) or a Wallastone prism, into two different beams travelling along different optical paths, one with horizontal and the other with vertical polarisation. Both beams impinge onto a polarisation-independent Digital Micromirror Device (DMD, DLP Light Crafter 6500 from Texas Instruments), at slightly different angles ($\approx 1.5^\circ$) and at the geometric centre of the DMD, where a digital hologram is displayed. The digital hologram contains the two constituting wave fields $\phi_1(x,y)$ and $\phi_2(x,y)$ required to generate the vector mode given by Eq. \ref{a42}. Each field is superimposed with a unique linear phase grating and both multiplexed into a single hologram. The period of each linear grating is carefully adjusted to ensure the overlap of the two first diffraction orders along a common propagation axis after the DMD, where the desired complex vector mode is generated. Afterwards, such mode is isolated from other diffraction orders by means of a Spatial Filter (SF), located at the focusing plane of a telescope formed by lenses $\rm L_3$ and $\rm L_4$, both with focal lengths $f=100$ mm. Finally, a Quarter-Wave plate (QW) is added to change the vector mode from the linear to the circular polarisation basis.

The generated modes where analysed in the {\bf Measuremet} stage through the use of a Circular Polariser (CP) fabricated by cementing a QWP to a linear polariser with its axis at $45^{\circ}$ with respect to the fast axis of the QWP. Here, the generated modes were focused onto the CCD camera with a lens of focal length $f=400$ mm. Transverse polarisation distributions were reconstructed using Stokes polarimetry, through a series of intensity measurements in accordance with the relations~\cite{Rosales2021}
\begin{equation}\label{Eq:Stokes}
\begin{split}
\centering
     &S_{0}=I_{H}+I_{V},\hspace{19mm}
     S_{1}=2I_{H}-S_{0},\\
     &S_{2}=2I_{D}-S_{0},\hspace{19mm}
     S_{3}=2I_{R}-S_{0},
\end{split}
\end{equation} 
where, $S_0$, $S_1$, $S_2$ and $S_3$ are the Stokes parameters and $I_H$, $I_V$, $I_D$ and $I_R$ represent the intensity of the horizontal, vertical, diagonal and right-handed polarisation components, respectively. Experimentally, such intensities where recorded with a Charge-Coupled Device (CCD) camera (FLIR FL3-U3-120S3C-C from Point Grey) in the {\bf Measurement} stage of Fig.\,\ref{setup} with the help of a CP. More specifically, $I_H$, $I_V$ and $I_D$ where obtained by setting the CP to $0^{\circ}$, $90^{\circ}$ and $45^{\circ}$, respectively, whereas $I_R$ was obtained by flipping the CP $180^{\circ}$ with its angle fixed to  $0^{\circ}$ (see for example \cite{Rosales2020,Rosales2021} for further details).

%%%%%%%%%%%%%%%%%%%%%%%%%%%%%%%%%%%%%%%%
\subsection{Transverse polarisation distribution in spatially disjoint vector fields}
\begin{figure*}[htb]
    \centering
    \includegraphics[width=0.9\textwidth]{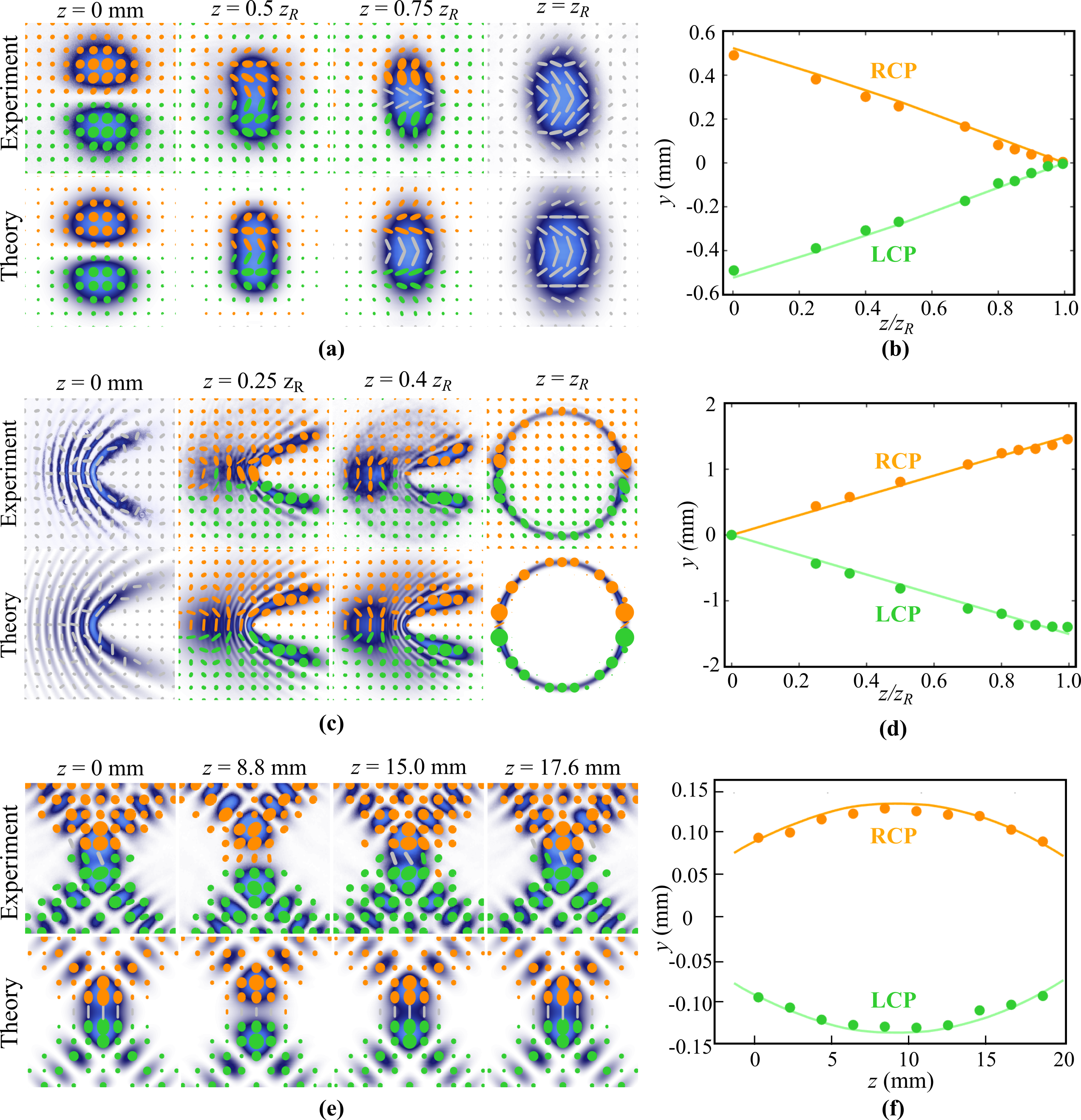}
    \caption{Examples of spatially disjoint vector fields. (a), (c) and (e) illustrates the transverse polarisation distribution and intensity profile at different propagation distances for Hermite-, Parabolic- and Airy-Gauss vector modes. (b), (d) and (f) illustrate the vertical separation of both polarisation components.}
    \label{Stokes}
\end{figure*}

In this section we will show three specific cases of engineered vector beams, whose polarisation structure changes dramatically upon free space propagation, namely, Hermite-, Parabolic- and Airy-Gauss vector modes. The left panels of Fig.\,\ref{Stokes} show the transverse polarisation distribution overlapped on the intensity profile of such modes for different propagation distances, both experimentally and theoretically. The right ones show the position of the centroid of the right- and left-handed circular polarisation components as function of the propagation distance.

%%%%%%%%%%%%%%%%%%%%
\subsubsection{First case: Hermite-Gauss vector beams}
As a first case, we analyse the example mentioned in the theory section, constructed from the non-separable superposition of the HG$_{01}$ mode, whose spatial profile features two vertical lobes of maximum intensity. Such mode was engineered in such a way that the upper lobe carries right-handed circular polarisation, while the lower one carries the orthogonal left-handed one. This mode is then focused with a lens of focal distance $f=400$ mm and scanned with the CCD camera along its propagation axis. The transverse polarisation distribution of such modes was reconstructed at selected planes $z=0$, $z=0.5z_R$, $z=0.75z_R$ and $z=z_R$, where $z_R$ is the Rayleigh length. Such polarisation distributions overlapped on the spatial shape of the modes are shown in Fig.\,\ref{Stokes}(a), experiment on the top row and numerical simulations, performed using the Rayleigh–Sommerfeld diffraction theory \cite{Goodman2017}, on the bottom. Fig.\,\ref{Stokes} (b) further illustrates how both modes approach to each other while their polarisation distribution evolves from quasi-homogeneous to non homogeneous, as seen in the approaching centroids of left- and right-handed polarisation components upon propagation.

%%%%%%%%%%%%%%%%%%%%
\subsubsection{Second case: Parabolic-Gauss beams}
As a second example, we analyse the recently introduced Parabolic-Gauss vector modes, which is implemented from a non-separable superposition of orthogonal parabolic modes, natural solutions to the Helmholtz equation in parabolic cylindrical coordinates, and orthogonal left- and right-handed polarisation states \cite{XiaoboHu2021}. Such vector modes feature an interesting behaviour as function of their propagation in free space, namely, their degree of non-separability evolves  from a non-homogeneously polarised vector beam to a quasi-homogeneously polarised one. Figure\,\ref{Stokes} (c) illustrates this behaviour for different propagation distances $z=0$, $z=0.25z_R$, $z=0.4z_R$ and $z=z_R$. Here we show the intensity distribution overlapped on the transverse polarisation distribution, experiment on top and theory on the bottom, which we reconstructed through Stokes polarimetry. Notice that for $z=0$ the transverse polarisation distribution contains only linear polarisation states whereas for $z=z_R=f$, the mode mainly contains left- and right-handed polarisation states. In Fig.\,\ref{Stokes} (d) we show the centroid of each of the circular polarisation components as function of their propagation distance. In this case, the separation increases from zero to a maximum separation, which is attained in the far field or the focal plane of the Fourier transforming lens ($z=f$).

%%%%%%%%%%%%%%%%%%%%
\subsubsection{Third case: Airy-Gauss beams}
As a final example, we engineered a vector mode consisting on the non-separable superposition of two Airy-Gauss beams with in-plane opposite transverse accelerations and orthogonal polarisation. As it is well-known, Airy-Gauss beams propagate in free space following a parabolic trajectory, whose parameters can be easily controlled \cite{Hu2010}. We superimposed two Airy-Gauss beams with orthogonal circular polarisation with their maximum intensity lobes facing each other. The polarisation and intensity distribution of such modes, reconstructed for specific planes, is illustrated in Fig.\,\ref{Stokes} (e), experiment on top and numerical simulations on the bottom. As illustrated, upon propagation the upper mode with right-handed circular polarisation moves upwards to a maximum point and then downwards. Similarly, the lower mode move downwards to a minimum point and then upwards. As result, upon propagation the modes initially separate from each other up to a maximum value, after which they approach to each other again. The trajectory of the main lobe of each beam as function of the propagation distance is shown in Fig.\,\ref{Stokes} (f). Here, we choose arbitrarily $z=0$ as the plane where the two modes are the closest to each other.

%%%%%%%%%%%%%%%%%%%%%%%%%%%%%%%%%%%%%%%%
\subsection{Determining the non-separability of spatially disjoint vector fields}

\begin{figure*}[htb]
    \centering
    \includegraphics[width=0.98\textwidth]{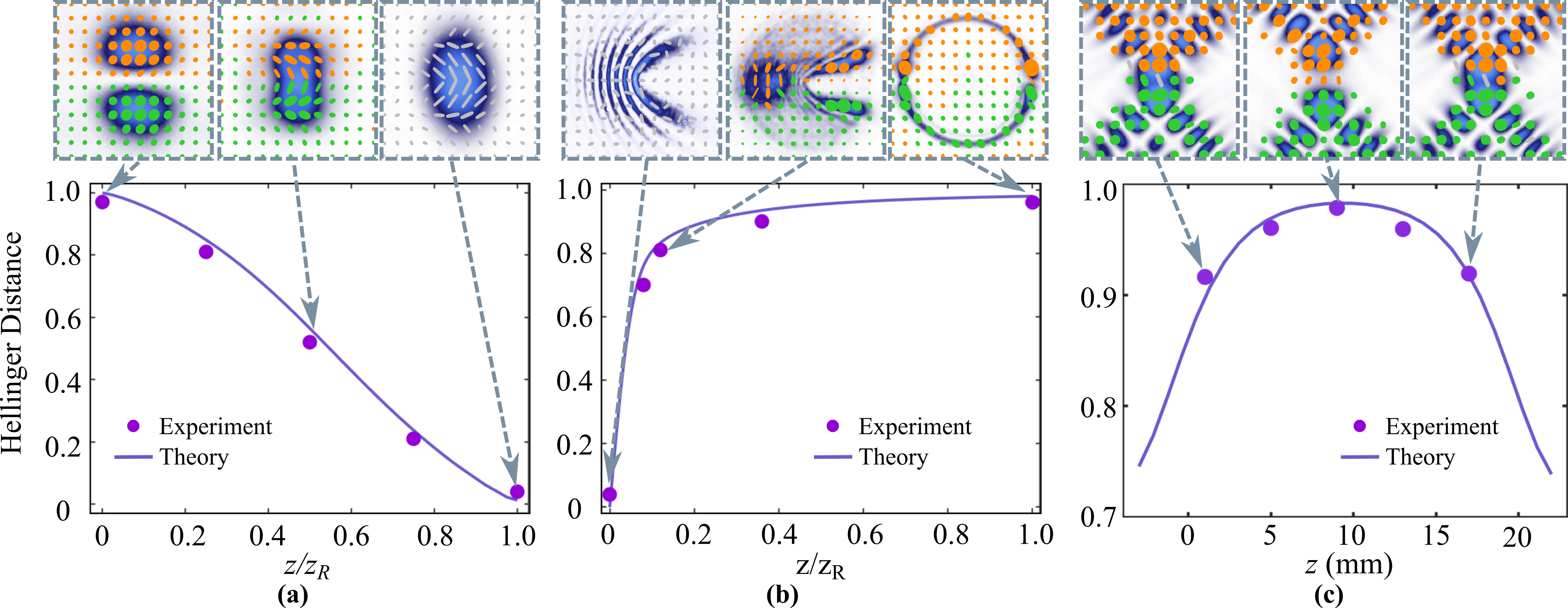}
    \caption{Hellinger distance as function of the propagation distance $z$ for (a) Hermite-, (b) Parabolic- and (c) Airy-Gauss vector modes.  The top insets show the experimentally reconstructed transverse polarisation distribution at the planes indicated by the arrows.}
    \label{Hdist}
\end{figure*}

As stated earlier, the degree of concurrence or non-separability of spatially disjoint vector fields, cannot be quantified using traditional methods, such as a basis independent tomography, as exhibited in \cite{Selyem2019}. In other words, traditional methods can only quantify the global behaviour of vector fields, but not the local behaviour exhibited by spatially disjoint vector fields. Hence, we introduced the Hellinger distance  (Eq. \ref{hd90}) as a measure that quantifies the local evolution of the non-separability in spatially disjoint vector fields. To demonstrate this, we applied Eq. \ref{hd90} to the three vector fields introduced in the previous section. Figure\,\ref{Hdist} (a) shows the case of the Hermite-Gauss vector field, where the Hellinger distance is plotted as function of propagation. For the sake of clarity, the transverse polarisation distribution at three different planes are shown as insets on the top of this figure, with arrows indicating their $z$ value. As mentioned in the theory, the Hellinger distance, which evolves from 1 to 0, clearly captures the evolution of the mode, from two modes with orthogonal circular polarisation into a single mode with linear polarisation. The second case corresponds to the parabolic-Gauss vector modes shown in Fig.\,\ref{Hdist} (b). Here, the transverse polarisation distribution evolves upon propagation from a single mode with linear polarisation to two spatially disjoint modes with circular orthogonal polarisation. Again, this  behaviour is captured by the Hellinger distance, which in this case increases monotonically from 0 to 1. The final example corresponds to the Airy-Gauss vector modes shown in Fig.\,\ref{Hdist} (c). This case was specifically engineered to show the effect of a parabolic-like Hellinger distance directly related to the transverse positions of the independent Airy-Gauss modes. In all cases, the theoretical results are shown as solid lines and the experimental data as solid circles, featuring remarkable agreement.

%%%%%%%%%%%%%%%%%%%%%%%%%%%%%%%%%%%%%%%%
\section{Discussion}  % Andrew
%%%%%%%%%%%%%%%%%%%%%%%%%%%%%%%%%%%%%%%%
Control over light's degrees of freedom is steadily gaining traction \cite{shen2021rays}, driven by the explosion of applications that structured light affords.  Concomitantly, there is a need to understand and quantify such new forms of light.  In the early 1990s, exotic scalar fields became possible by the emergence of diffractive optical elements, which saw the measurement toolkit expanded to include statistical approaches (intensity moments), resulting in (for example) the beam quality factor, $M^2$.  This single parameter is not sufficient to fully describe such light fields, and thus the toolkit quickly matured, forming the basis of our present ISO standards for laser beams.  More recently, vectorially structured light is routinely created and employed in a variety of applications, but analogous to the situation for scalar light 30 years ago, our toolkit is very much in its infancy.  This time our toolkit is quantum-inspired, returning a vector quality factor (the vectorial equivalent to the $M^2$ parameter), but with some shortcomings that require continual improvement.  The first was to alter the measurement from basis dependent \cite{McLaren2015} to a basis independent \cite{Selyem2019}, removing the requirement of knowledge of the field to be probed. But here we have highlighted that even though these techniques perform very well at characterising vector modes whose spatial and polarisation degrees of freedom are shape-invariant upon propagation, it fails at characterising beams that do not satisfy such condition, whose constituting spatial modes separate from each other upon propagation, becoming path-like separable. The existing approaches are insufficient to quantify such disjoint behaviour. We envisaged a way in which we could address this issue by defining a Hellinger distance, between the constituting spatial modes that form the vector beam, as a measure. We hope that this work inspires further advances in the toolkit.
%(i) redefine the local concurrence so that there is no arbitrariness in choosing the region for its computation, and (ii) define

Our work also raises some interesting fundamental questions on the nature of visibility (spatial coherence) in the context of vectorially structured light.  Our propagation induced disjointness is analogous to a double slit experiment, but as if the interference pattern is time reversed to return to the double slits.  The question is how the visibility changes with propagation? In our analogous situation, the intial locally inhomogeneous field, with high visibility in the polarisation variation, evolves to become locally homogeneous with no visibility - we can observe (intuitively) two distinct regions made of scalar fields (our two ``slits'').  Yet the global field retains its visibility during propagation, as if the original slits themselves are now the interference pattern with high visibility (1 inside the slits and 0 outside).  

The situation has a quantum analogue too.  While it is commonplace in quantum optics to exploit the path as a degree of freedom for entanglement, this is largely untouched in the context of classical quantum-like light \cite{pabon2020high,shen2021creation,shen2020structured}.  When the paths are marked, no interference is observed. In our disjoint field, this is equivalent to measuring locally, one scalar component.  When the paths are not marked, full visibility is possible.  In our classical field, this is equivalent to making projections on the entire field, a global concurrence measurement. The physics of path embedded information in such vectorial light fields thus requires further probing, and will surely shed light on extending notions of spatial coherence to multi-partite, multiple degree of freedom light \cite{qian2020quantification}.

%%%%%%%%%%%%%%%%%%%%%%%%%%%%%%%%%%%%%%%%
\section{Conclusions}  % Andrew
%%%%%%%%%%%%%%%%%%%%%%%%%%%%%%%%%%%%%%%%
Here we have revisited the emerging field of quantum-like vectorial (classical) light, and highlighted the shortcomings of traditional quantum-inspired measures. We considered the general case where the local polarisation homogeneity can vary spatially across the field, from scalar to vector, by virtue of a propagation induced spatial disjointness.  We proposed and demonstrated a new measure that correctly accounts for the local separability of the field, while accounting for the fact that the global vectorial field remains non-separable. Our work is relevant fundamentally, probing the notion of non-separability and spatial coherence in the context of vectorially polarised light, but also practically, where new standards to universally quantify these fields are a pressing need. 
%%%%%%%%%%%%%%%%%%%%%%%%%%%%%%%%%%%%%%%%
%%%%%%%%%%%%%%%%%%%%%%%%%%%%%%%%%%%%%%%%
\section{Acknowledgements}  
AA acknowledges support from the Deutsche Forschungsgemeinschaft Project No. 429529648-
TRR 306 QuCoLiMa (``Quantum Cooperativity of Light and Matter''). CRG acknowledges support from the National Natural Science Foundation of China (61975047 and 11934013) and the High-Level Talents Project of Heilongjiang Province (Grant No. 2020GSP12)

\section*{References}
%\bibliography{References}

\appendix

\section{Hermite-Gauss modes}\label{HGmodes}

The two-dimensional Hermite-Gauss modes $u_{nm}(\bm{x},z) = \varphi_n(x,z) \varphi_m(y,z)$, with $n,m= 0,1,\ldots, \infty$, are defined by \cite{Siegman},
\begin{align}\label{aa110}
\varphi_{n}(x,z) = & \; \frac{1}{\pi^{1/4}} \frac{1}{ \sqrt{2^n n!}} \, \frac{1}{\sqrt{\rho}} \, \text{H}_n \! \left( { x}/{\rho} \right) e^{-(x/\rho)^2/2} \nonumber \\[4pt]
& \times  e^{\frac{i}{2}\frac{z}{z_R} (x/\rho)^2} \, e^{ - i \left( n + \frac{1}{2} \right) \arctan \left( {z}/{z_R} \right) }.
\end{align}
In Eq. \eqref{aa110} $\text{H}_n(x)$ denotes the  $n$th-order Hermite polynomial, and
\begin{align}\label{aa120}
\rho = \rho(z) =  \frac{w_0}{\sqrt{2}}  \left( 1+ \frac{z^2}{z_R^2} \right)^{1/2},
\end{align}
fixes the transverse length scale at distance $z$ from the beam's origin, where the minimum beam radius   $w_0 > 0$, is attained. The Rayleigh length $z_R =k w_0^2/ 2$,  sets the longitudinal length scale, giving the distance over which the beam can propagate without spreading significantly. From Eqs. \eqref{aa110} and  \eqref{aa120} it readily follows that
\begin{align}\label{aa220}
u_{nm}(\bm{x},z) = \frac{1}{\xi(z)} u_{nm}\left( \frac{\bm{x}}{\xi(z)},0 \right)  \exp \bigl[ i \chi_{nm}(\bm{x},z)\bigr],
\end{align}
with
\begin{align}\label{aa130}
\xi(z) = \frac{\rho(z)}{\rho(0)} = \left(1+\frac{z^2}{z_R^2} \right)^{1/2},
\end{align}
and
\begin{align}\label{aa230}
\chi_{nm}(\bm{x},z) = & \;  \frac{z}{z_R} \left| \frac{1}{w_0} \frac{\bm{x}}{\xi(z)} \right|^2 \nonumber \\[6pt]
& -  (n+m+1) \arctan \left( \frac{z}{z_R} \right).
\end{align}

\section{Quantum-like notation}\label{notation}

In this appendix we quickly review the common use of a quantum-like notation for the study of vector beams.
To begin with, let $\bm{\Psi}(\bm{x},z)$ be the vector amplitude of a non-uniformly polarized paraxial optical beam propagating along the $z$-axis of a Cartesian reference frame with coordinates $(x,y,z) = (\bm{x},z)$, defined by
\begin{align}\label{aa10}
\bm{\Psi}(\bm{x},z) = \hat{\bm{e}}_1 \, \psi_1(\bm{x},z) + \hat{\bm{e}}_2 \, \psi_2(\bm{x},z),
\end{align}
where the real-valued unit vectors $\hat{\bm{e}}_1$ and $\hat{\bm{e}}_2$ are parallel to the $x$- and the $y$-axis, respectively, with $\hat{\bm{e}}_1 \cdot \hat{\bm{e}}_2 = 0$, and $\psi_1(\bm{x},z),\psi_2(\bm{x},z)$ are two arbitrary solutions of the paraxial wave equation, which do not need to be orthogonal and normalized. The \emph{global} intensity of the beam is denoted $I_0$, and defined by
\begin{align}\label{aa20}
I_0 = & \;  \intR \bm{\Psi}^*(\bm{x},z) \cdot \bm{\Psi}(\bm{x},z) \, {\text{d}^2 x} \nonumber \\[8pt]
= & \;  \intR \Bigl\{ \abs{\psi_1(\bm{x},z)}^2 + \abs{\psi_2(\bm{x},z)}^2 \Bigr\} \, {\text{d}^2 x},
\end{align}
which is independent of $z$.

Now, we introduce the \emph{discrete} ``polarisation basis'' $\{\ket{1} , \ket{2} \}$, and the \emph{continuous} ``position basis'' $\{\ket{ \bm{x}}  = \ket{x,y} \}$, with $\bm{x} \in \mathbb{R}^2$.
By definition, the polarisation basis $\{ \ket{i} \}_{i \in \{1,2\}}$ spans the discrete \emph{polarisation Hilbert space} $\mathcal{H}_\text{pol}$, and the position basis $\{ \ket{\bm{x}} \}_{\bm{x} \in \mathbb{R}^2}$  spans the continuous \emph{position Hilbert space} $\mathcal{H}_\text{pos}$.
We postulate that these bases are orthonormal
\begin{align}
\brak{\alpha}{\beta} = & \;  \delta_{\alpha \beta},  \phantom{xxxxxxxxxxxxxx} (\alpha, \beta =1,2), \label{aa30A} \\[8pt]
\brak{\bm{x}}{\bm{x}'} = & \; \brak{x}{x'}\brak{y}{y'} \nonumber \\[8pt]
 = & \; \delta(x - x') \delta(y - y')\nonumber \\[8pt]
\equiv & \; \delta(\bm{x} - \bm{x}') , \phantom{xxxxxxxxxx} (\bm{x}, \bm{x}' \in \mathbb{R}^2), \label{aa30B}
\end{align}
and complete:
\begin{align}
\sum_{\alpha=1}^2 \proj{\alpha}{\alpha} & \; = \hat{I}_\text{pol}, \label{aa40A} \\[8pt]
\int_{\mathbb{R}^2} \proj{\bm{x}}{\bm{x}} \, {\text{d}^2 x} & \; = \hat{I}_\text{pos}, \label{aa40B}
\end{align}
where $\hat{I}_\text{pol}$ and $\hat{I}_\text{pos}$ are the identity operators in the polarisation and position spaces, respectively.
By definition, the direct product $\ket{\alpha} \otimes \ket{\bm{x}} = \ket{\alpha}\ket{\bm{x}} = \ket{\alpha,\bm{x}}$ spans the whole Hilbert space  $\mathcal{H} = \mathcal{H}_\text{pol} \otimes \mathcal{H}_\text{pos}$.

Using this quantum-like notation (but the physics of the problem is entirely classical), we can represent $\bm{\Psi}(\bm{x},z)$ by means of the state vector $\ket{\Psi(z)}$ defined by
\begin{align}\label{aa50}
\ket{\Psi(z)} = & \; \ket{1} \ket{ \psi_1(z)} + \ket{2} \ket{ \psi_2(z)}\nonumber \\[8pt]
= & \; \sum_{\beta =1}^{2} \ket{\beta} \ket{ \psi_\beta(z)},
\end{align}
such that
\begin{align}\label{aa60}
\brak{\alpha, \bm{x}}{\Psi(z)} = & \; \sum_{\beta=1}^{2} \brak{\alpha}{\beta} \brak{\bm{x}}{ \psi_\beta(z)} \nonumber \\[8pt]
= & \; \brak{\bm{x}}{ \psi_\alpha(z)}\nonumber \\[8pt]
= & \; \psi_\alpha(\bm{x},z) , \qquad (\alpha=1,2),
\end{align}
where $\psi_1(\bm{x},z),\psi_2(\bm{x},z)$ are defined by \eqref{aa10}.
For the sake of clarity,  here and hereafter we shall occasionally write  $\ket{ \psi_\alpha}$ for $\ket{ \psi_\alpha(z)}$ and $\ket{ \Psi}$ for $\ket{ \Psi(z)}$, the $z$-dependence being understood. Moreover, we shall use upper case Greek letters, as $ \Psi$, to denote vectors in   $\mathcal{H}$, while lower case Greek letters, as $ \psi$, will denote vectors in   $\mathcal{H}_\text{pos}$.

Using this notation, we can rewrite the intensity of the beam as
\begin{align}\label{aa70}
I_0 = & \; \intR \Bigl\{ \abs{\psi_1(\bm{x},z)}^2 + \abs{\psi_2(\bm{x},z)}^2 \Bigr\} {\text{d}^2 x} \nonumber \\[8pt]
= & \; \brak{ \psi_1 }{ \psi_1 } + \brak{ \psi_2 }{ \psi_2 } \nonumber \\[8pt]
= & \; \brak{ \Psi }{ \Psi },
\end{align}
where, here and hereafter, for any $\ket{ \psi }, \ket{ \phi } \in \mathcal{H}_\text{pos}$, and $\ket{ \Psi } \in \mathcal{H}$, we write
\begin{align}\label{aa80}
\brak{ \psi }{ \phi } = & \; \bra{ \psi } \hat{I}_\text{pos} \ket{ \phi } \nonumber \\[8pt]
= & \; \intR \brak{ \psi }{\bm{x}} \brak{\bm{x}}{ \phi } \, {\text{d}^2 x}  \nonumber \\[8pt]
= & \; \intR  \psi^* (\bm{x},z) \phi(\bm{x},z) \, {\text{d}^2 x},
\end{align}
and
\begin{align}\label{aa90}
\brak{ \Psi }{ \Psi } = & \; \bra{ \Psi } \hat{I}_\text{pol} \otimes \hat{I}_\text{pos} \ket{ \Psi } \nonumber \\[8pt]
= & \; \sum_{\alpha=1}^2 \intR \brak{ \Psi}{\alpha,\bm{x}} \brak{\alpha,\bm{x}}{ \Psi} \, {\text{d}^2 x} \nonumber \\[8pt]
= & \;  \sum_{\alpha=1}^2 \intR \brak{ \psi_\alpha }{\bm{x}} \brak{\bm{x}}{ \psi_\alpha } \, {\text{d}^2 x}  \nonumber \\[8pt]
= & \;  \sum_{\alpha=1}^2 \brak{ \psi_\alpha }{ \psi_\alpha },
\end{align}
where \eqref{aa60} has been used.

\section{The Schmidt decomposition}\label{appSchmidt}

Given the field
\begin{align}\label{e10}
\bm{\Psi}(\bm{x},z) = \hat{\bm{e}}_1 \, \psi_1(\bm{x},z) + \hat{\bm{e}}_2 \, \psi_2(\bm{x},z),
\end{align}
and the corresponding state vector
\begin{align}\label{e20}
\ket{\Psi(z)} = \ket{1} \ket{ \psi_1(z)} + \ket{2} \ket{ \psi_2(z)},
\end{align}
the Schmidt decomposition is easy to evaluate.
First, using $\ket{\psi_1}$ and $\ket{\psi_2}$ we build the orthonormal basis $\{\ket{v_1},\ket{v_2}\}$ defined by
\begin{align}\label{c240}
\ket{v_1} = \frac{\ket{\psi_1}}{\sqrt{\brak{\psi_1}{\psi_1}}} ,
\end{align}
and
\begin{align}\label{c250}
\ket{v_2} = & \; \frac{\ket{\psi_2} - \ket{v_1}\brak{v_1}{\psi_2}}{\sqrt{\brak{\psi_2}{\psi_2} - \abs{\brak{v_1}{\psi_2}}^2}}\nonumber \\[8pt]
= & \; \frac{\displaystyle \ket{\psi_2} - \ket{\psi_1}\frac{\displaystyle \brak{\psi_1}{\psi_2}}{\brak{\psi_1}{\psi_1}}}{\sqrt{\brak{\psi_2}{\psi_2} - \frac{\displaystyle \abs{\brak{\psi_1}{\psi_2}}^2}{\displaystyle \brak{\psi_1}{\psi_1}}}}.
\end{align}
Equations \eqref{c240} and \eqref{c250} imply
\begin{align}\label{c260}
\ket{\psi_1} = \ket{v_1} \sqrt{\brak{\psi_1}{\psi_1}} \, ,
\end{align}
and
\begin{align}\label{c270}
\ket{\psi_2} = \ket{v_1}\frac{\brak{\psi_1}{\psi_2}}{\sqrt{\brak{\psi_1}{\psi_1}}} + \ket{v_2} \sqrt{\brak{\psi_2}{\psi_2} - \frac{\displaystyle \abs{\brak{\psi_1}{\psi_2}}^2}{\displaystyle \brak{\psi_1}{\psi_1}}}.
\end{align}
Substituting (\ref{c260}-\ref{c270}) into \eqref{e20} we find
\begin{align}\label{c280}
\ket{\Psi(z)} = & \; \ket{1}  \,\ket{v_1} \left(  \sqrt{\brak{\psi_1}{\psi_1}} \right) + \ket{2}  \, \ket{v_1}\left(   \,
\frac{\brak{\psi_1}{\psi_2}}{\sqrt{\brak{\psi_1}{\psi_1}}} \right) \nonumber \\[8pt]
& + \ket{2}  \, \ket{v_2} \left(   \, \sqrt{\brak{\psi_2}{\psi_2} - \frac{\displaystyle \abs{\brak{\psi_1}{\psi_2}}^2}{\displaystyle \brak{\psi_1}{\psi_1}}} \right) \nonumber \\[8pt]
\equiv & \; \sum_{\alpha, \beta =1}^2 A_{\alpha \beta} \, \ket{\alpha}  \ket{v_\beta}.
\end{align}
The $2 \times 2$ matrix $A$ is defined by \eqref{c280} as
\begin{align}\label{c290}
A = \begin{bmatrix}
       \sqrt{\brak{\psi_1}{\psi_1}} & 0 \\[8pt]
        \,
\frac{\displaystyle \brak{\psi_1}{\psi_2}}{\displaystyle \sqrt{\brak{\psi_1}{\psi_1}}} &   \, \sqrt{\displaystyle \brak{\psi_2}{\psi_2} - \frac{\displaystyle \abs{\brak{\psi_1}{\psi_2}}^2}{\displaystyle \brak{\psi_1}{\psi_1}}} \\[8pt]
    \end{bmatrix},
\end{align}
and  it is independent of $z$. Substituting the singular value decomposition
\begin{align}\label{c300}
A = U D V^\dagger,
\end{align}
into \eqref{c280}, we can easily calculate the Schmidt form of $\ket{\Psi(z)}$ as:
\begin{align}\label{c310}
\ket{\Psi(z)} = & \; \sum_{\alpha =1}^2 D_\alpha \left( \sum_{\beta = 1}^2 U_{\beta \alpha} \ket{\beta} \right) \left( \sum_{\gamma =1}^2 V^\dagger_{\alpha \gamma} \ket{v_\gamma} \right)\nonumber \\[8pt]
\equiv & \; \sum_{\alpha =1}^2 \sqrt{\lambda_\alpha}  \, \ket{\varepsilon_\alpha} \ket{\phi_\alpha},
\end{align}
where $D_{\alpha \beta} = \delta_{\alpha \beta} D_\alpha \equiv  \delta_{\alpha \beta} \sqrt{\lambda_\alpha}$, $\lambda_\alpha$ being the non-negative eigenvalues of $A^\dagger A$, and we have defined
\begin{equation}\label{c315}
\begin{split}
\ket{\varepsilon_\alpha} \equiv & \;  \sum_{\beta = 1}^2 U_{\beta \alpha} \ket{\beta}, \\[8pt]
\ket{\phi_\alpha} \equiv & \; \sum_{\gamma =1}^2 V^\dagger_{\alpha \gamma} \ket{v_\gamma} .
\end{split}
\end{equation}
A straightforward calculation gives
\begin{equation}\label{eigen}
\begin{split}
\lambda_1 = & \;  \lambda_+, \\[8pt]
\lambda_2 = & \;  \lambda_-,
\end{split}
\end{equation}
where
\begin{align}
\lambda_\pm = & \;  \frac{1}{2} \biggl[ \brak{\psi_1}{\psi_1} + \brak{\psi_2}{\psi_2} \nonumber \\[8pt]
 & \phantom{ \frac{1}{2} \biggl[} \pm \sqrt{4 \abs{\brak{\psi_1}{\psi_2}}^2 + \left( \brak{\psi_1}{\psi_1} - \brak{\psi_2}{\psi_2} \right)^2} \biggr].
\end{align}
Note that since the state vector $\ket{\Psi(z)}$ is not normalized, then
\begin{align}\label{c320}
\lambda_1 + \lambda_2 = \brak{\psi_1}{\psi_1}   + \brak{\psi_2}{\psi_2}   \neq 1.
\end{align}
Finally, multiplying  \eqref{c310} from left by $\bra{\bm{x}}$ and using Eqs. \eqref{aa50} and \eqref{aa60}, we obtain the Schmidt form of the field \eqref{e10}:
\begin{align}\label{a100bis}
\bm{\Psi}(\bm{x},z) = \sqrt{\lambda_1} \, \hat{\bm{\varepsilon}}_1 \, \phi_1(\bm{x},z) + \sqrt{\lambda_2} \, \hat{\bm{\varepsilon}}_2 \, \phi_2(\bm{x},z).
\end{align}

The amount  of the entanglement in the vector state $\ket{\Psi(z)}$  can be quantified by the so-called Schmidt number, denoted $K$ and defined by
\begin{align}\label{c330}
K = & \; \frac{\left(\lambda_1 + \lambda_2 \right)^2}{\lambda_1^2 + \lambda_2^2}\nonumber \\[8pt]
= & \; \frac{\displaystyle  \bigl( \brak{\psi_1}{\psi_1}   + \brak{\psi_2}{\psi_2}   \bigr)^2}{\displaystyle \brak{\psi_1}{\psi_1}^2   + 2 \abs{\brak{\psi_1}{\psi_2}}^2     + \brak{\psi_2}{\psi_2}^2  } \, .
\end{align}
For a bipartite $2$-dimensional system the Schmidt number $K$ and the concurrence $C$ are simply related:
\begin{subequations}\label{c486}
\begin{align}
K = & \;  \frac{2}{2- C^2} \, , \label{c486A} \\[8pt]
C = & \; \sqrt{2 \left(1-\frac{1}{K}\right)} = 2 \,\frac{ \sqrt{\lambda_1 \lambda_2}}{\lambda_1 + \lambda_2} \, . \label{c486B}
\end{align}
\end{subequations}
We present an explicit calculation of $C$ in Appendix \ref{quattro}.

\section{Density matrix operator and coherency matrix}\label{tre}

The density matrix operator $\hat{\rho}$ for the pure state $\ket{\Psi}$ is defined as the projector
\begin{align}\label{l80}
\hat{\rho} = \frac{\proj{\Psi}{\Psi}}{\brak{ \Psi }{ \Psi }}.
\end{align}
The reduced polarisation density matrix operator $\hat{\rho}_\text{pol}$, is calculated by tracing $\hat{\rho}$ with respect to the position degrees of freedom, as follows:
\begin{align}\label{l90}
\hat{\rho}_\text{pol} = & \; \intR \mean{\bm{x}}{\hat{\rho}}{\bm{x}} \, {\text{d}^2 x} \nonumber \\[8pt]
= & \; \frac{1}{\brak{ \Psi }{ \Psi }} \intR \brak{\bm{x}}{\Psi} \brak{\Psi}{\bm{x}} \, {\text{d}^2 x} \nonumber \\[8pt]
= & \; \frac{1}{\brak{ \Psi }{ \Psi }} \sum_{\alpha,\beta=1}^2 \proj{\alpha}{\beta} \intR \brak{ \psi_\beta}{\bm{x}} \brak{\bm{x}}{ \psi_\alpha} \, {\text{d}^2 x} \nonumber \\[8pt]
= & \; \frac{1}{\brak{ \Psi }{ \Psi }} \sum_{\alpha,\beta=1}^2 \proj{\alpha}{\beta}  \brak{ \psi_\beta}{ \psi_\alpha}.
\end{align}
It is easy to check that $\tr \hat{\rho}_\text{pol} =1$. This equations shows that in the polarisation basis $\{ \ket{1} , \ket{2} \}$, $\hat{\rho}_\text{pol}$ is represented by the $2 \times 2$  normalized coherency matrix of the beam, denoted $\rho_\text{pol}  = J/ \tr J$, and defined by
\begin{align}\label{l100}
\hat{\rho}_\text{pol}  \doteq & \; \rho_\text{pol} \nonumber \\[8pt]
= & \; \begin{bmatrix}
         \mean{1}{\hat{\rho}_\text{pol}}{1} & \mean{1}{\hat{\rho}_\text{pol}}{2} \\[4pt]
         \mean{2}{\hat{\rho}_\text{pol}}{1} & \mean{2}{\hat{\rho}_\text{pol}}{2} \\
       \end{bmatrix}
\nonumber \\[8pt]
= & \; \frac{1}{\brak{ \Psi }{ \Psi }} \begin{bmatrix}
         \brak{\psi_1}{\psi_1} & \brak{\psi_2}{\psi_1} \\[4pt]
         \brak{\psi_1}{\psi_2} & \brak{\psi_2}{\psi_2} \\
       \end{bmatrix}\nonumber \\[8pt]
= & \; \frac{1}{\tr J} \, J \,,
\end{align}
where $\tr J = \brak{ \Psi }{ \Psi } = \brak{\psi_1}{\psi_1} +  \brak{\psi_2}{\psi_2}$. A direct calculation shows that
\begin{align}\label{similar}
J = A  A^\dagger ,
\end{align}
 where $A$ is defined by Eq. \eqref{c290}. This implies that $J$ and $A^\dagger A$ have the same eigenvalues $\lambda_1,\lambda_2$ given in \eqref{eigen}.

The coherency matrix $J$ can be written in terms of the so-called Stokes parameters $\{S_0,S_1,S_2,S_3 \}$ as,
\begin{align}\label{l110}
J = & \; \frac{1}{2} \begin{bmatrix}
         S_0 + S_1 & S_2 - i S_3 \\[4pt]
         S_2 + i S_3 & S_0 - S_1 \\
       \end{bmatrix}.
\end{align}
Comparing \eqref{l100} and \eqref{l110} we obtain
\begin{align}\label{l120}
\frac{1}{2} \begin{bmatrix}
         S_0 + S_1 & S_2 - i S_3 \\[4pt]
         S_2 + i S_3 & S_0 - S_1 \\
       \end{bmatrix} =
\begin{bmatrix}
         \brak{\psi_1}{\psi_1} & \brak{\psi_2}{\psi_1} \\[4pt]
         \brak{\psi_1}{\psi_2} & \brak{\psi_2}{\psi_2} \\
       \end{bmatrix}.
\end{align}
This evidently implies
\begin{equation}\label{l140}
\begin{split}
\frac{S_0 + S_1}{2} = & \; \brak{\psi_1}{\psi_1},  \\[8pt]
\frac{S_0 - S_1}{2} = & \;\brak{\psi_2}{\psi_2},  \\[8pt]
\frac{S_2 - i S_3}{2} = & \;\brak{\psi_2}{\psi_1}, \\[8pt]
\frac{S_2 + i S_3}{2} = & \; \brak{\psi_1}{\psi_2}.
\end{split}
\end{equation}
Solving this set of four equations with respect to $\{S_0,S_1,S_2,S_3 \}$, we find
\begin{equation}\label{l130}
\begin{split}
S_0 = & \; \brak{\psi_1}{\psi_1} + \brak{\psi_2}{\psi_2},  \\[8pt]
S_1 = & \; \brak{\psi_1}{\psi_1} - \brak{\psi_2}{\psi_2},  \\[8pt]
S_2 = & \; \brak{\psi_1}{\psi_2} + \brak{\psi_2}{\psi_1} , \\[8pt]
S_3 = & \; \frac{\brak{\psi_1}{\psi_2} - \brak{\psi_2}{\psi_1}}{i} ,
\end{split}
\end{equation}
so that we can rewrite \eqref{l100} as
\begin{align}\label{l135}
\rho_\text{pol} = & \; \frac{1}{ \brak{\psi_1}{\psi_1} +  \brak{\psi_2}{\psi_2}} \begin{bmatrix}
         \brak{\psi_1}{\psi_1} & \brak{\psi_2}{\psi_1} \\[4pt]
         \brak{\psi_1}{\psi_2} & \brak{\psi_2}{\psi_2} \\
       \end{bmatrix} \nonumber \\[8pt]
= & \;\frac{1}{2 S_0} \begin{bmatrix}
         S_0 + S_1 & S_2 - i S_3 \\[4pt]
         S_2 + i S_3 & S_0 - S_1 \\
       \end{bmatrix}.
\end{align}

\section{The concurrence {\it C}}\label{quattro}

The concurrence $C$ of the state vector  $\ket{\Psi}$ is defined as \cite{PhysRevA.71.012318},
\begin{align}\label{l150}
C =  \sqrt{2 \left(1 - \tr \rho_\text{pol}^2  \right) },
\end{align}
where
\begin{align}\label{l160}
\tr \rho_\text{pol}^2 = & \; \frac{\tr \left\{ \begin{bmatrix}
         \brak{\psi_1}{\psi_1} & \brak{\psi_2}{\psi_1} \\[4pt]
         \brak{\psi_1}{\psi_2} & \brak{\psi_2}{\psi_2} \\
       \end{bmatrix}^2  \right\} }{\brak{\Psi}{\Psi}^2} \nonumber \\[8pt]
  = & \; \frac{ \brak{\psi_1}{\psi_1}^2 + 2 \brak{\psi_2}{\psi_1}\brak{\psi_1}{\psi_2} + \brak{\psi_2}{\psi_2}^2 }{\left( \brak{\psi_1}{\psi_1} +  \brak{\psi_2}{\psi_2} \right)^2} \nonumber \\[8pt]
  = & \; \frac{ \left(S_0 + S_1 \right)^2 + 2 \abs{S_2 - i S_3 }^2 + \left(S_0 - S_1 \right)^2 }{4 S_0^2} \nonumber \\[8pt]
  = & \; \frac{1}{2} + \frac{S_1^2 + S_2^2 + S_3^2}{2 S_0^2},
\end{align}
and Eq. \eqref{l140} has been used. Substituting \eqref{l160} into \eqref{l150} we obtain
\begin{align}\label{l170}
C = \sqrt{1 - \frac{S_1^2 + S_2^2 + S_3^2}{S_0^2} } \,.
\end{align}

Using the second line of \eqref{l160} and $\brak{ \Psi }{ \Psi } = \brak{\psi_1}{\psi_1} +  \brak{\psi_2}{\psi_2}$, we can also write the concurrence as
\begin{align}\label{new10}
C =  2 \, \frac{\sqrt{\vphantom{\bigl[} \brak{\psi_2}{\psi_2}  \brak{\psi_1}{\psi_1} - \brak{\psi_2}{\psi_1}\brak{\psi_1}{\psi_2} }}{\brak{\psi_1}{\psi_1} +  \brak{\psi_2}{\psi_2}}.
\end{align}
Then we remember that the Schmidt decomposition of $\ket{\Psi}$ is (see Appendix \ref{appSchmidt}),
\begin{align}\label{new20}
\ket{\Psi} = & \; \sqrt{\lambda_1} \, \ket{\varepsilon_1} \ket{ \phi_1} +  \sqrt{\lambda_2} \, \ket{\varepsilon_2} \ket{ \phi_2} \nonumber \\[8pt]
\equiv  & \;  \ket{\varepsilon_1} \ket{ \psi_1} +  \ket{\varepsilon_2} \ket{ \psi_2},
\end{align}
where
\begin{align}\label{new30}
\brak{\varepsilon_\alpha}{\varepsilon_\beta} = \delta_{\alpha \beta}, \qquad \text{and} \qquad \brak{\phi_\alpha}{\phi_\beta} = \delta_{\alpha \beta},
\end{align}
with $\alpha,\beta =1,2$.
Substituting \eqref{new20} into \eqref{new10} we obtain
\begin{align}\label{new40}
C = & \;  2 \, \frac{\sqrt{\vphantom{\prod}  \lambda_1 \lambda_2}}{\lambda_1+\lambda_2} \sqrt{ \vphantom{\bigl[}  \brak{\phi_2}{\phi_2}  \brak{\phi_1}{\phi_1} - \brak{\phi_2}{\phi_1}\brak{\phi_1}{\phi_2} }\nonumber \\[8pt]
=  & \; 2 \, \frac{\sqrt{\vphantom{\prod} \lambda_1 \lambda_2}}{\lambda_1+\lambda_2} ,
\end{align}
where \eqref{new30} has been used.

As a consistency check, we repeat now the same calculation for $C$, but using the equivalent definition
\begin{align}\label{l190}
C =  \sqrt{2 \left( 1 - \tr \hat{\rho}_\text{pos}^2 \right) }\, ,
\end{align}
where the reduced position density matrix operator $\hat{\rho}_\text{pos}$ is calculated by tracing $\hat{\rho}$ with respect to the polarisation degrees of freedom,
\begin{align}\label{l180}
\hat{\rho}_\text{pos} = & \; \sum_{\alpha=1}^2 \mean{\alpha}{\hat{\rho}}{\alpha} \nonumber \\[8pt]
= & \; \frac{1}{\brak{ \Psi }{ \Psi }} \sum_{\alpha=1}^2 \brak{\alpha}{\Psi} \brak{\Psi}{\alpha} \nonumber \\[8pt]
= & \; \frac{1}{\brak{ \Psi }{ \Psi }} \Bigl(
 \proj{\psi_1}{\psi_1} +  \proj{\psi_2}{\psi_2}  \Bigr).
\end{align}
Note that this quantity is \emph{not} a $2 \times 2$ matrix, but an infinite-dimensional operator, as it can be directly seen by rewriting $\hat{\rho}_\text{pos}$ in the position basis:
\begin{align}\label{l185}
\hat{\rho}_\text{pos} = & \; \hat{I}_\text{pos} \, \hat{\rho}_\text{pos} \, \hat{I}_\text{pos} \nonumber \\[8pt]
= & \; \intR {\text{d}^2 x} \intR {\text{d}^2 x'} \, \ket{\bm{x}} \mean{\bm{x}}{ \hat{\rho}_\text{pos}}{\bm{x}'} \bra{\bm{x}'} \nonumber \\[8pt]
= & \; \frac{1}{\brak{ \Psi }{ \Psi }} \intR {\text{d}^2 x} \intR {\text{d}^2 x'} \, \proj{\bm{x}}{\bm{x}'} \nonumber \\[8pt]
& \times\Bigl[ \psi_1(\bm{x},z)\psi_1^*(\bm{x}',z)
+\psi_2(\bm{x},z)\psi_2^*(\bm{x}',z)  \Bigr] .
\end{align}
From this equation it follows that
\begin{align}\label{l200}
\tr \hat{\rho}_\text{pos}^2 = & \; \frac{1}{\brak{\Psi}{\Psi}^2} \tr \left[ \left(
 \sum_{\alpha = 1}^2 \proj{\psi_\alpha}{\psi_\alpha}  \right) \left(\sum_{\beta = 1}^2 \proj{\psi_\beta}{\psi_\beta}  \right) \right] \nonumber \\[8pt]
 = & \; \frac{1}{\brak{\Psi}{\Psi}^2} \sum_{\alpha,\beta =1}^2 \brak{\psi_\alpha}{\psi_\beta} \tr \bigl(
 \proj{\psi_\alpha}{\psi_\beta} \bigr)  \nonumber \\[8pt]
  = & \; \frac{1}{\brak{\Psi}{\Psi}^2} \sum_{\alpha,\beta =1}^2 \abs{\brak{\psi_\alpha}{\psi_\beta}}^2 \nonumber \\[8pt]
  = & \; \frac{1}{\brak{\Psi}{\Psi}^2} \Bigl[
 \brak{\psi_1}{\psi_1}^2 + 2 \abs{\brak{\psi_1}{\psi_2}}^2    +  \brak{\psi_2}{\psi_2}^2 \Bigr] \nonumber \\[8pt]
  = & \; \frac{1}{2} + \frac{S_1^2 + S_2^2 + S_3^2}{2 S_0^2},
\end{align}
where \eqref{l140} has been used, and
\begin{align}\label{l210}
\tr \bigl( \proj{\psi_\alpha}{\psi_\beta} \bigr) = & \; \intR \brak{\bm{x}}{\psi_\alpha} \brak{\psi_\beta}{\bm{x}} \, {\text{d}^2 x} \nonumber \\[8pt]
= & \; \brak{\psi_\beta}{\psi_\alpha}, \qquad \qquad \qquad (\alpha,\beta = 1,2).
\end{align}
 Substituting \eqref{l200} into \eqref{l190} we obtain
\begin{align}\label{l220}
C = \sqrt{1 - \frac{S_1^2 + S_2^2 + S_3^2}{S_0^2} } \,.
\end{align}
As expected, we have obtained the same value for the concurrence $C$ by using either \eqref{l150} or \eqref{l190}.
\end{document}